\documentclass[%
prl,
reprint,
superscriptaddress,
amsmath,
amssymb,
aps,
]{revtex4-2}

\usepackage{graphicx}
\usepackage{bm}
\usepackage{amsmath,amsfonts,amssymb}
\usepackage{physics}
\usepackage{hyperref}
\usepackage{url}
\usepackage{txfonts}
\usepackage{dsfont}
\usepackage{mathtools}

\usepackage{tikz}
\usepackage{pgfplots}
\usetikzlibrary{decorations.pathmorphing, positioning, shapes,arrows,arrows.meta}
\usetikzlibrary{shapes.geometric}
\usetikzlibrary{fadings}
\usetikzlibrary{external}
\usepgfplotslibrary{fillbetween}
\tikzfading 
[
  name=fade out,
  inner color=transparent!0,
  outer color=transparent!100
]
\usepackage{natbib}

\usetikzlibrary{external}
\begin{document}
\title{Quantum limits for resolving Gaussian sources}
\author{Giacomo Sorelli} 
\affiliation{Laboratoire Kastler Brossel, Sorbonne Universit\'e, ENS-Universit\'e  PSL, CNRS, Collège de France, 4  Place Jussieu, F-75252 Paris, France}
\author{Manuel Gessner}
\affiliation{ICFO-Institut de Ci\`{e}ncies Fot\`{o}niques, The Barcelona Institute of Science and Technology, Av. Carl Friedrich Gauss 3, 08860, Castelldefels (Barcelona), Spain}
\author{Mattia Walschaers} 
\affiliation{Laboratoire Kastler Brossel, Sorbonne Universit\'e, ENS-Universit\'e PSL, CNRS, Collège de France, 4  Place Jussieu, F-75252 Paris, France}
\author{Nicolas Treps}
\affiliation{Laboratoire Kastler Brossel, Sorbonne Universit\'e, ENS-Universit\'e PSL, CNRS, Collège de France, 4  Place Jussieu, F-75252 Paris, France}

\date{\today}
\begin{abstract}
We determine analytically the quantum Cram\'er-Rao bound for the estimation of the separation between two point sources in arbitrary Gaussian states.
Our analytical expression is valid for arbitrary sources brightness, and it allows to determine how different resources, such as mutual coherence (induced by thermal correlations or displacement) or squeezing affect the scaling of the ultimate resolution limit with the mean number of emitted photons.
In practical scenarios, we find coherent states of the sources to achieve quantum optimal resolution. 
\end{abstract}
\maketitle

{\it Introduction.---}
Resolving two point sources, i.e. establishing their separation with diffraction-limited imaging systems becomes more difficult the closer the sources are. 
This intuitive statement lies at the basis of historical resolution criteria \cite{Abbe,Rayleigh}, which established minimal distances beyond which diffraction renders it impossible to resolve two sources.  
However, in the last decades a great number of {\it superresolution} techniques allowed to resolve separations beyond the diffraction limit, by either intervening on the properties of the sources \cite{Hell:94,Klar8206,Betzig1642} (active imaging) or of the measurements \cite{Helstrom73,Hsu_2004,Delaubert_2008,Tsang_review} (passive imaging).

The possibility to overcome the diffraction limit inspired the search for ultimate resolution limits. 
The natural framework for this investigation is that of quantum metrology \cite{helstrom1976,holevo2011probabilistic,GiovannettiLoydMaccone,Paris2009,Luca_Augusto_review}.
The latter establishes that the quantum limit on the estimation of the source separation $d$ is given by the quantum Cram\'er-Rao bound $(\Delta d)^2 \geq F_d^{-1}$, with $F_d$ the quantum Fisher information (QFI) that quantifies the sensitivity of the quantum state of the sources to variations of the separation $d$.
The quantum estimation approach showed that for two dim incoherent point sources the QFI is independent of their separation \cite{Tsang_PRX}. Accordingly, in this case, when optimal measurements are performed, arbitrary separations can be resolved with the same sensitivity.
Moreover, it was proved theoretically \cite{Tsang_PRX,Rehacek:17} and verified experimentally \cite{Paur:16,Tang:16,Yang:16,Tham:2017,Boucher:20,Zanforlin:2022}, that this ultimate resolution limit can be achieved by spatial mode demultiplexing followed by intensity measurements. 
These results have been extended to incoherent thermal sources \cite{Nair_2016} and other photon-number diagonal states \cite{LupoPirandola}. 
More recently, the role of partial coherence 
has been widely discussed \cite{Larson:18, tsang2019comment, Larson19reply, Hradil19QFI, Hradil21Partial, kurdzialek2021sources,Liang:21,Wadood:21,De:2021,Tsang2021poissonquantum}.

Despite this vast body of research works, ultimate resolution limits are known for a very limited class of quantum states of the sources. 
Furthermore, with the notable exceptions of the bounds in \cite{Nair_2016, LupoPirandola}, all known results are valid only for low photon fluxes in the image plane.
In this letter, we overcome these limitations presenting an analytical expression for the QFI for the estimation of the separation between two point sources in arbitrary Gaussian states. 
Such states include fully coherent and incoherent (thermal) states, but also those (quantum or classically) correlated states that are most widely accessible in experiments \cite{Holevo:1975,Weedbrook:2012,Adesso:2014}.
Accordingly, the associated resolution limits are practically relevant for both passive and active imaging.

Studying Gaussian states, we have access to different types of mutual coherence between the sources, and we can evaluate their impact beyond the low flux limit.
In this high brightness regime, we are able to continuously interpolate between fully mutually coherent and thermal sources, and to show that mutual coherence originating from thermal correlations and displacement are non-equivalent. 
They lead to different scaling of the QFI with the mean photon number emitted by the sources, with displacement always achieving the best performances.
Accordingly,  coherent states of the sources enable a resolution that surpasses that of every classically correlated thermal state.
In fact, we show that, in all practically relevant scenarios, coherent states, with an optimized phase relation, approach the ultimate achievable resolution, as given in \cite{LupoPirandola}.


{\it Imaging Gaussian sources.--- }
We consider two point sources, that are transversally separated by a distance $d$ in the object plane, and are observed through a diffraction-limited imaging system with a real point spread function (PSF) $u_0({\bf r})$.
Without loss of generality, we assume that the sources are aligned along the $x-$axis,
and populate the two localized source modes at $\pm {\bf r}_0 = (\pm d/2,0)$ with associated quadratures $\hat{\bf x}^{(s)} = (\hat{q}_1^{(s)}, \hat{p}_1^{(s)},\hat{q}_2^{(s)},\hat{p}_2^{(s)})$ satisfying the commutation relations $[\hat{q}^{(s)}_j,\hat{p}^{(s)}_k] = 2i\delta_{jk}$, and $[\hat{q}^{(s)}_j,\hat{q}^{(s)}_k] = [\hat{p}^{(s)}_j,\hat{p}^{(s)}_k] =0$.
We restrict ourselves to Gaussian states $\rho$ of the sources, which are fully determined by the first two moments of the quadratures \cite{Holevo:1975,Weedbrook:2012,Adesso:2014}: the mean field $\bar{\bf x}^{(s)} = \tr [ \hat{x}^{(s)}\rho]$, and the covariance matrix $V^{(s)}_{jk} = \tr \left[ \left\{\hat{x}^{(s)}_j - \bar{x}^{(s)}_j, \hat{x}^{(s)}_k - \bar{x}^{(s)}_k\right\} \rho \right]/2$, with $\{\cdot, \cdot\}$ denoting the anti-commutator.
\begin{figure}[t]
    \centering
    \includegraphics[width=\columnwidth]{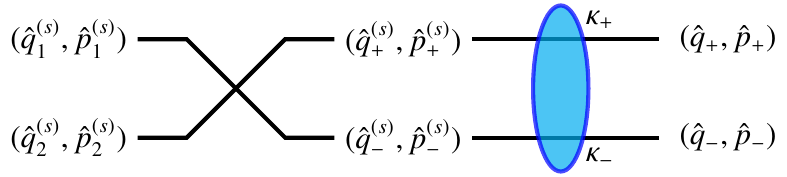}
    \caption{Graphical illustration of the propagation model: quadrature operators, $(\hat{q}_{1,2}^{(s)},\hat{p}_{1,2}^{(s)})$, of the localized source modes are transformed into their symmetric and antisymmetric superpositions $(\hat{q}_\pm^{(s)},\hat{p}_\pm^{(s)})$. The latter evolve through independent loss channels, with transmissivities $\kappa_\pm$ into the image-plane quadrature operators $(\hat{q}_\pm,\hat{p}_\pm)$.}
    \label{Fig:model}
    \end{figure}
    
In the far field, the imaging system is characterized by a transmission coefficient $\kappa \ll 1$, and maps the localized sources modes into the overlapping modes $u_0({\bf r} + {\bf r}_0)$, (with associated quadratures $\hat{q}_1$ and $\hat{p}_1$) and $u_0({\bf r} - {\bf r}_0)$, (with associated quadratures $\hat{q}_2$ and $\hat{p}_2$).
The diffraction-induced non-orthogonality of the modes $u_0({\bf r} \pm {\bf r}_0)$ induces the non-commutativity of the associated quadratures: $[\hat{q}_1, \hat{p}_2] = [\hat{q}_2, \hat{p}_1] = 2 i \delta$, with $\delta = \int  u_0({\bf r} - {\bf r}_0)u_0({\bf r} + {\bf r}_0) d {\bf r}$.
We therefore introduce the orthonormal image modes $u_\pm({\bf r}) = [u_0({\bf r} - {\bf r}_0) \pm u_0({\bf r} + {\bf r}_0)]/\sqrt{2(1\pm\delta)}$ with associated quadratures 
\begin{equation}
    \hat{q}_\pm = \frac{\hat{q}_1 \pm \hat{q}_2}{\sqrt{2(1\pm\delta)}}, \quad \hat{p}_\pm = \frac{\hat{p}_1 \pm \hat{p}_2}{\sqrt{2(1\pm\delta)}}.
    \label{qp_pm}
\end{equation}
As illustrated in Fig.~\ref{Fig:model}, the image-plane quadratures $\hat{q}_\pm$ are related to those of the symmetric and antisymmetric superpositions of the localized sources modes $\hat{q}^{(s)}_\pm$ according to $\hat{q}_\pm= \sqrt{\kappa_\pm}\hat{q}^{(s)}_\pm + \sqrt{1-\kappa_\pm}\hat{q}^{(v)}_\pm$ (and analogously for $\hat{p}_\pm$), with $(\hat{q}^{(v)}_\pm,\hat{p}^{(v)}_\pm)$ the quadratures associated with two auxiliary vacuum modes, and $\kappa_\pm = (1 \pm \delta)\kappa$ are parameter-dependent transmission coefficients of two independent loss channels \cite{LupoPirandola}. 
Therefore, Gaussian states of the sources evolve through the imaging system into Gaussian states of the orthonormal image modes $u_\pm({\bf r})$. 
The covariance matrix $V$ and the mean field $\bar{\bf x}$ of such image-plane states are connected with those in the localized source modes according to $\bar{\bf x} = T \bar{\bf x}^{(s)}$ and $V = T V^{(s)} T^\top+ N$, with 
\begin{align}
    T =
    \begin{pmatrix}
    \sqrt{\frac{\kappa_+}{2}} \mathds{1}_2 & \sqrt{\frac{\kappa_+}{2}} \mathds{1}_2 \\
     \sqrt{\frac{\kappa_-}{2}}\mathds{1}_2 & - \sqrt{\frac{\kappa_-}{2}} \mathds{1}_2
    \end{pmatrix}, \;
     N =  \begin{pmatrix}
     (1 - \kappa_+)\mathds{1}_2 & 0 \\
     0 & (1 - \kappa_-)\mathds{1}_2\\
     \end{pmatrix}.
\end{align}

The state of the sources in the image plane depends on the separation $d$ through the shape of the populated modes $u_\pm({\bf r})$, and, because of the parameter-dependent transmission coefficients $\kappa_\pm$, through the mean field $\bar{\bf x}$ and covariance matrix $V$.
Aided by the Williamson decomposition of the image-plane covariance matrix, $V = S (\nu_+ \mathds{1}_2 \oplus \nu_- \mathds{1}_2) S^\top$, with symplectic eigenvalues $\nu_\pm$, and $S$ a symplectic matrix \cite{Weedbrook:2012,Adesso:2014}, we account for these different dependencies and compute the QFI $F_d$ for the estimation of the separation $d$ \cite{Sorelli:2022}:
\begin{subequations}
\begin{align}
F_d &= F_V + F_{\bar{\bf x}}, \quad {\rm with}\\
F_V &=\frac{1}{2} \sum_{l=0}^3\sum_{jk=\pm} \left[\frac{\left(a_{jk}^{(l)}\right)^2}{\nu_j\nu_k- (-1)^l} +2\frac{\left(\tilde{a}_{k,j}^{(l)}\right)^2}{\nu_j- (-1)^l}\right], \label{QFI_covariance} \\ 
F_{\bar{\bf x}} &= (\partial_d \bar{\bf x})^\top V^{-1} (\partial_d \bar{\bf x})+\bar{\bf x}^\top D_\partial^2 \bar{\bf x},
\label{QFI_displacement}
\end{align}
\label{QFI}
\end{subequations}
where $a^{(l)}_{ij} = \tr \left[ A^{(l)}_{ij} S^{-1} (\partial_d V) S\right]$, $\tilde{a}^{(l)}_{ij} = \tr \left[A^{(l)}_{ij}  S^{-1}(V - \mathds{1}_4) D_\partial \right]$, and $D_\partial = \eta_+ \mathds{1}_2 \oplus \eta_- \mathds{1}_2$ with $\eta^2_\pm = \int |\partial_d u_\pm ({\bf r}) |^2 d {\bf r}$, and $A^{(l)}_{jk}$ are a basis of the space of $4 \times 4$ matrices \footnote{The matrices $A^{(l)}_{jk}$ are zero everywhere except in the $jk$ block where they are given by $i\sigma_y/\sqrt{2}$ (for $l=0$), $\sigma_z/\sqrt{2}$ (for $l=1$), $\mathds{1}_2/\sqrt{2}$ (for $l=2$) and $\sigma_x/\sqrt{2}$ (for $l=3$), with $\sigma_{j=x,y,z}$ the standard Pauli matrices}.

Let us now comment on the physical origin of the different terms of the QFI~\ref{QFI}:
$F_{\bar{\bf x}}$~\eqref{QFI_displacement} describes how the sensitivity is affected by changes of the mean field $\bar{\bf x}$ in the modes $u_\pm({\bf r})$.  
In particular, the first term in Eq.~\eqref{QFI_displacement} contains the derivative $\partial_d \bar{\bf x}$ which account for mean field variations due to parameter-dependent losses, as quantified by the transmission coefficients $\kappa_\pm$.
The second term in Eq.~\eqref{QFI_displacement} contains the diagonal matrix $D_\partial$ whose elements $\eta_\pm$ quantify how much the shape of the modes $u_\pm({\bf r})$ changes with the separation $d$.

In a similar fashion, $F_V$~\eqref{QFI_covariance} describes the sensitivity due to variations of the covariance matrix $V$ with $d$.
More specifically, Eq.~\eqref{QFI_covariance} contains a sum running over the components $a_{jk}$ and $\tilde{a}_{jk}$ (on the basis $A_{jk}^{(l)}$) of the matrices $S^{-1} (\partial_d V) S$ and $S^{-1}(V - \mathds{1}_4) D_\partial$, respectively.
The former matrix contains the derivative $\partial_d V$, which describes changes of the covariance matrix due to the parameter-dependent transmissions $\kappa_\pm$, while the latter matrix contains $D_\partial$, which, as discussed above, takes into account the $d-$dependence of the modes $u_\pm({\bf r})$.
In the following, we discuss various examples to illustrate the impact of these different contributions on our ability to resolve two point sources. 

{\it Mutual coherence.---}
\begin{figure}[t]
    \centering
      \includegraphics[width=\columnwidth]{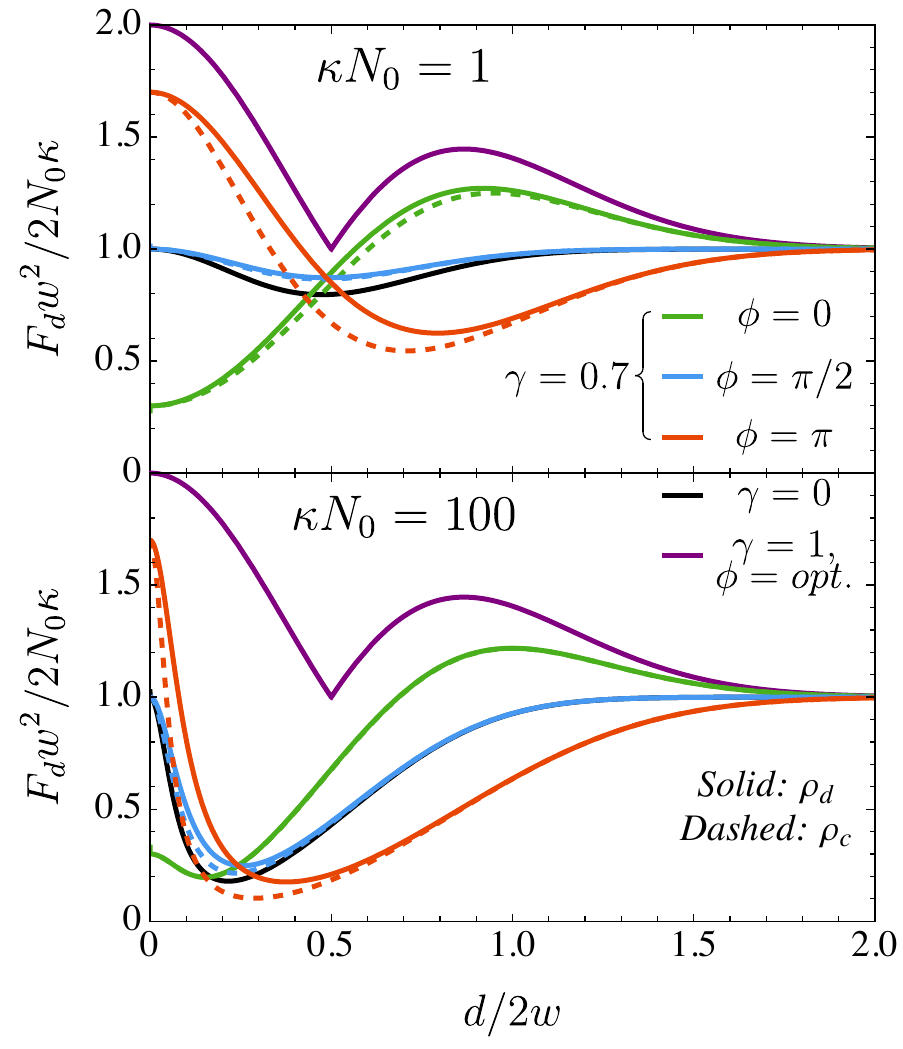}
    \caption{Quantum Fisher information $F_d$ for two partially mutually coherent sources ($\gamma = 0.7$), with coherence originating from thermal correlations (dashed lines) or displacement (solid lines) as a function of their transverse separation $d$.
    Top and bottom panels correspond to $\kappa N_0 = 1$ and  $\kappa N_0 = 100$, respectively. 
    Different colors represent different phases of the degree of mutual coherence ($\phi = 0$ in green, $\phi = \pi/e$ in blue, $\phi = \pi$ in red).
    Black and purple curves correspond to the mutual incoherent case ($\gamma =0$), and to coherent sources with an optimized mutual phase $(\phi =0, \pi)$ depending on $d$), respectively.}
    \label{fig:partial_coherence}
\end{figure}
In optics, two light sources are said to be (partially) mutually coherent, if they present a definite phase relationship that allows them to interfere \cite{mandel_wolf,klauder_sudarshan}.
This capacity to interfere can be quantified, in quantum mechanical terms, through the complex degree of mutual coherence (DMC)
\begin{equation}
   \gamma e^{i\phi} =  \frac{\langle \hat{s}^\dagger_1 \hat{s}_2 \rangle}{\sqrt{\langle \hat{s}^\dagger_1 \hat{s}_1 \rangle \langle \hat{s}^\dagger_2 \hat{s}_2 \rangle}}.
   \label{DMC}
\end{equation}
The amplitude of the DMC ranges from $\gamma = 0$ for mutually incoherent, i.e. uncorrelated thermal,  sources, to $\gamma = 1$ in the case of perfect mutual coherence.
On the other hand, the phase in Eq.~\eqref{DMC} distinguishes different kinds of interference, with $\phi =0$ ($\phi =\pm \pi$) corresponding to perfectly constructive (destructive) interference.

In this work, we focus on two different states of two sources, each emitting on average $N_0$ photons, and featuring (partial) mutual coherence. 
First, we consider correlated thermal states $\rho_{c}$, which are zero mean Gaussian states, with covariance matrix 
\begin{equation}
    V^{(s)} =
\begin{pmatrix}
(2N_0 +1) \mathds{1}_2 & 2N_0\gamma R(\phi) \\
- 2N_0\gamma R(\phi) & (2N_0 +1) \mathds{1}_2
\end{pmatrix}
\end{equation}
with $R(\phi)$ a rotation matrix. 
Second, we consider the two sources to be in identical uncorrelated thermal states equally displaced along two directions separated by an angle $\phi$ in phase space.  
These displaced thermal states $\rho_{d}$ have a diagonal covariance matrix $ V^{(s)} = [2N_0(1-\gamma) +1] \mathds{1}_4$, and a mean field $\bar{\bf x}^{(s)} = 2\sqrt{\gamma N_0}(1, 0,\cos \phi, \sin \phi)$ \footnote{Here, without loss of generality, we assumed the first source to be displaced along the $q-$axis of phase space.}.
Both $\rho_c$ and $\rho_d$ have a DMC $\gamma e^{i\phi}$. 
However, for $\rho_c$, $\gamma$ corresponds to the strength of the thermal correlations, while $\phi$ is their relative phase.
On the other hand, for $\rho_d$, $\gamma$ is the displacement amplitude (in unit of the mean photon number), and $\phi$ is the relative phase between the two displacements. 
The QFI~\eqref{QFI}, for correlated and displaced thermal states, is presented in Fig.~\ref{fig:partial_coherence}, for an imaging system with Gaussian PSF $u_0({\bf r}) = \sqrt{2/(\pi w^2)}\exp (-|{\bf r}|^2/w^2)$.

On the one hand, $\rho_c$ and $\rho_d$ produce similar dependencies of the QFI on the source separation: $F_d \to 2N_0\kappa/w^2$ for distances larger than the PSF width ($d \gg w$), and $ F_{d\to 0} = 2\kappa N_0(1-\gamma\cos\phi)/w^2$.
Accordingly, for small separations, when compared to the fully incoherent case ($\gamma =0$, black curves in Fig.~\ref{fig:partial_coherence}) partial coherence enhances the QFI in the case of destructive interference ($|\phi| > \pi/2$), and reduces it for constructive interference ($|\phi| < \pi/2$).

On the other hand, the QFI for displaced thermal sources (solid lines in Fig.~\ref{fig:partial_coherence}) is always larger than that for correlated thermal sources (dashed lines in Fig.~\ref{fig:partial_coherence}).  
In particular, for coherent states of the sources ($\rho_d$ with $\gamma =1$), we have 
\begin{equation}
    F_d = 2\kappa N_0\left((\Delta k)^2 - \beta \cos \phi \right),
    \label{QFI_coherent}
\end{equation}
where we have introduced
\begin{subequations}
\begin{align}
    (\Delta k)^2 &= \int [\partial_x u_0 ({\bf r})]^2 d {\bf r}, \\
    \beta &= \int \partial_x u_0 ({\bf r}-{\bf r}_0)\partial_x u_0 ({\bf r}+{\bf r}_0)d {\bf r}.
\end{align}
\end{subequations}
From Eq.~\eqref{QFI_coherent}, we see that the QFI per emitted photon $F_d/N_0$ does not depend on the mean photon number $N_0$. 
Conversely, for perfectly correlated thermal sources ($\gamma =1$), for intermediate separations of the order of the PSF width $w$, the QFI decreases when increasing $N_0$. 
Accordingly, for two sources in coherent states, the separation-estimation sensitivity presents a shot-noise scaling $\Delta d \sim N_0^{-1/2}$, while for perfectly correlated thermal sources ($\rho_c$ with $\gamma =1$) the scaling is less favorable. 

In the low photon flux regime ($\kappa N_0 \ll 1$), the role of partial coherence  has been recently debated \cite{Larson:18, tsang2019comment, Larson19reply, Hradil19QFI, Hradil21Partial, kurdzialek2021sources,Liang:21,Wadood:21,De:2021,Tsang2021poissonquantum}.
The debate originated from two different models for the state of the sources in the image plane. 
The first, proposed by Larson and Saleh in \cite{Larson:18}, describes the sources in the image plane as a single photon state, while the other, introduced by Tsang and Nair in \cite{tsang2019comment}, considers a mixture of a single photon and a vacuum contribution.
This vacuum contribution allows to describe losses induced by the imaging system, which, in presence of mutual coherence, lead to a separation-dependent mean photon number in the image plane. Such an additional dependence can increase the QFI \cite{tsang2019comment,kurdzialek2021sources,Tsang2021poissonquantum}.
However, this information is not available if the mean photon number $N_0$ emitted by the sources is unknown, and in this case one recovers the results of the single-photon model \cite{Wadood:21,karuseichyk2021}. The two classes of Gaussian states considered here both contain a vacuum contribution. 
In particular, correlated thermal states are photon-number diagonal and can be considered a generalization of the Tsang and Nair model \cite{tsang2019comment}, to which they reduce in the low flux regime. 
On the contrary, displaced thermal states are not diagonal in the photon number basis and, even for $N_0 \ll 1$, do not correspond to any partial coherence model that has been studied in the low flux regime.

{\it Squeezing and entanglement.---}
\begin{figure}[t]
    \centering
    \includegraphics[width = \columnwidth]{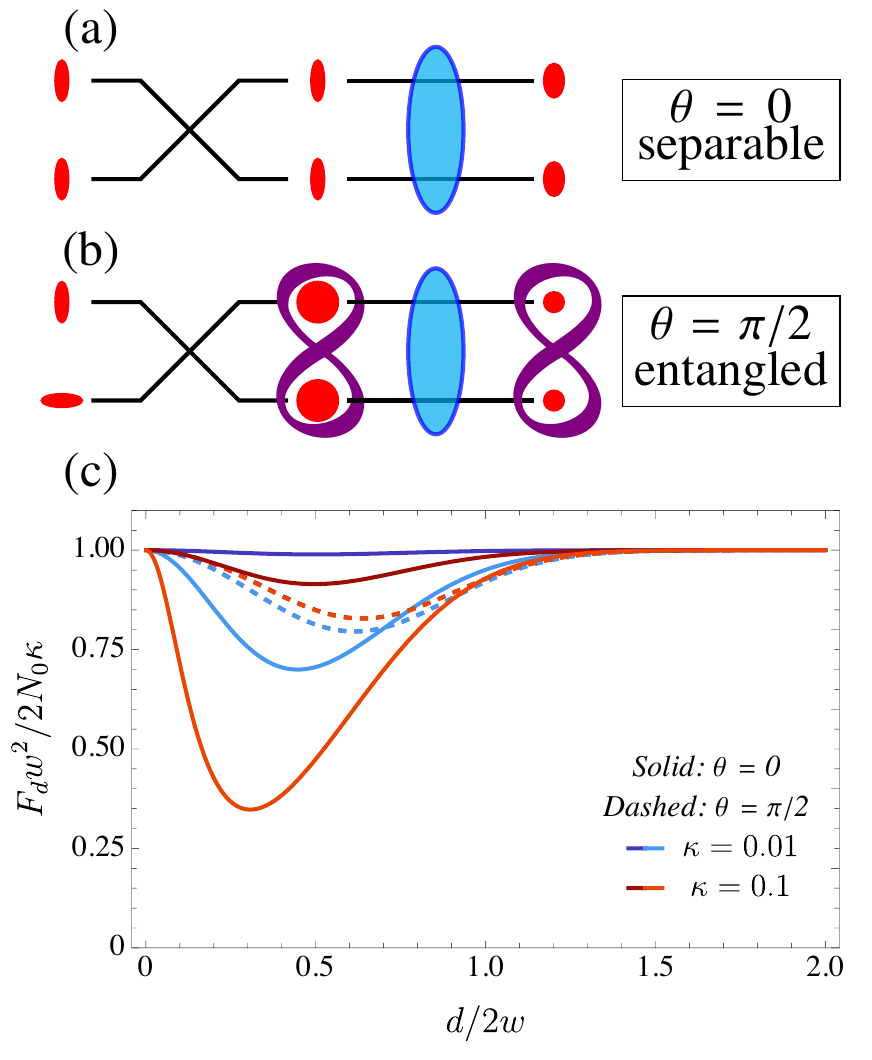}
    \caption{Graphical representation of the propagation of two sources squeezed along the same direction (a) and along opposite directions (b) to the image plane.
    (c) Quantum Fisher information $F_d$ as a function of their transverse separation $d$, for two sources squeezed along the same direction ($\theta =0$, solid lines) or along opposite directions ($\theta =\pi/2$, dashed lines) in phase space. 
    Different colors correspond to different transmission of the imaging system, $\kappa = 0.01$ (blue) and $\kappa = 0.1$ (red), and different shades to different mean photon numbers, $N_0=1$ (darker) and  $N_0=100$ (lighter).}
    \label{fig:squeezing}
\end{figure}
We have just seen that coherent states of the source modes provide the optimal resolution among mutual coherent, classically correlated, states.
We now investigate if quantum resources such as squeezing or entanglement can lead to enhanced performances. 
To this goal, we consider two zero-mean sources squeezed along two directions separated by an angle $\theta$ in phase space. 
Such sources have covariance matrix $V_0^\theta =S^2(\xi)\oplus R(\theta)S^2(\xi)R^\top(\theta)$,
with $S(\xi) = {\rm diag}(e^{-\xi},e^\xi)$ a single-mode squeezing matrix, and $R(\theta)$ a rotation matrix.
The mean photon number $N_0$ emitted by each source is related to the squeezing parameter $\xi$ according to $N_0 = \sinh^2 \xi$ \cite{Weedbrook:2012,serafini_book}.
In the following, we focus on two simple, but exemplary cases: two sources squeezed along the same direction ($\theta = 0$) and along opposite directions ($\theta = \pi/2$).

The evolution from the object to the image plane (see Fig.~\ref{Fig:model}) is given by a transformation to the symmetric and antisymmetric source modes, followed by two independent loss channels.
As illustrated in Fig.~\ref{fig:squeezing} (a), for $\theta = 0$, the mapping to the symmetric and antisymmetric source modes leaves the state unchanged, while propagation through the loss channels adds noise, such that the quantum state in the image modes becomes a product of two thermal squeezed states.
Similar results are obtained considering a two-mode squeezed vacuum of the localized source modes, corresponding to two vacuum states squeezed in opposite directions in the symmetric and antisymmetric source modes. The latter case was studied in the supplementary material of \cite{LupoPirandola}. 
On the contrary, for $\theta = \pi/2$, see Fig.~\ref{fig:squeezing} (b), we obtain a two-mode squeezed vacuum in the symmetric and antisymmetric source modes, which is entangled, and it remains entangled upon propagation to the image plane for every non-zero value of the source separation $d>0$.

The behaviour of the QFI for these two classes of squeezed states is presented in Fig.~\ref{fig:squeezing} (c). 
For sources squeezed along the same direction ($\theta =0$), the high propagation losses  ($\kappa \ll 1$ in the far field) 
render the state substantially semiclassical. 
Accordingly, their QFI (solid lines in Fig.~\ref{fig:squeezing} (c)) is very similar to that for uncorrelated thermal sources (black lines in Fig.~\ref{fig:partial_coherence}): it equals $F_d = 2N_0\kappa/w$ except for a region, around separations of the order of the PSF width $w$, where it presents a dip which gets deeper when increasing the mean number of photons $\kappa N_0$ in the image plane.
On the other hand, when the two sources are squeezed along opposite directions ($\theta = \pi/2$), despite the high losses, their state preserve its quantum character upon propagation through the imaging system, i.e. it remains entangled in the orthogonal image modes $u_\pm({\bf r})$.
As a consequence, in this case, the QFI per emitted photon $F_d/N_0$ is independent of the mean photon number $N_0$ (see dashed lines in Fig.~\ref{fig:squeezing} (c)).
Thus, entanglement enables a shot-noise scaling $\Delta d \sim N_0^{-1/2}$ of the separation estimation sensitivity. 
Contrarily, non-entangled squeezed states provide a less favourable scaling for intermediate separations.
Finally, comparing Fig.~\ref{fig:squeezing} (c) with Fig.~\ref{fig:partial_coherence}, we observe that several partially mutually coherent states allow for larger values of the QFI.
Thus, squeezing in the sources modes does not improve the separation estimation sensitivity with respect to semiclassical states.

\textit{Approaching the ultimate quantum limit.---}
Coherent sources ($\rho_d$ with $\gamma=1$) present the same shot-noise scaling $\Delta d \sim N_0^{-1/2}$ of the separation estimation sensitivity as entangled squeezed states, but with a more favourable prefactor. In fact, we can show that in all practically relevant scenarios, coherent states are quantum optimal:
Recent works demonstrated that, for the separation estimation sensitivity, it is impossible to achieve a sub-shot-noise scaling \cite{LupoPirandola,gessner2020}, and that for arbitrary quantum states of the sources the QFI cannot exceed the global upper bound \cite{LupoPirandola}
\begin{align}
    F_d^{\rm max} &= 2\kappa N_0\max \left\{f_+,f_-\right\},\quad {\rm with} \label{upper_bound}\\
    f_\pm &= (\Delta k)^2 \mp \beta +\frac{\kappa (\partial_d \delta)^2}{1-\kappa(1\pm\delta)}. \label{fpm}
\end{align}
For imaging in the far field regime ($\kappa \ll 1$), the last term in Eq.~\eqref{fpm} is negligible ($f_\pm \approx (\Delta k)^2 \mp \beta$), and therefore the global upper bound \eqref{upper_bound} is achieved by the QFI \eqref{QFI_coherent} for coherent sources either in or out of phase (purple curves in Fig.~\ref{fig:partial_coherence}).

{\it Conclusion.---}
We determined the ultimate quantum resolution limit for two point sources in arbitrary Gaussian states.
Such a limit is provided in the form of an analytical expression for the QFI quantifying the sensitivity of the quantum state of the sources to variations of their separation.
With this general expression for the QFI, we explored the the role of different (non-equivalent) types of partial coherence, and classical and quantum correlations in the two source resolution problem. 
Our results show that semi-classical states of the sources featuring (partial) coherence can often outperform sources featuring quantum properties such as squeezing. 
In fact, coherent states of the sources can even approach the ultimate quantum resolution, in the practically relevant far field regime.


{\it Acknowledgments.---}
This work was partially funded by French ANR under COSMIC project (ANR-19-ASTR0020-01). This work received funding from the European Union’s Horizon 2020 research and innovation programme under Grant Agreement No. 899587. This work
was supported by the European Union’s Horizon 2020 research and innovation programme under the QuantERA
programme through the project ApresSF. This work received funding from Ministerio de Ciencia e Innovaci\'{o}n (MCIN) / Agencia Estatal de Investigaci\'on (AEI) for Project No. PID2020-115761RJ-I00 and support of a fellowship from ``la Caixa” Foundation (ID 100010434) and from the European Union’s Horizon 2020 research and innovation program under Marie Sk\l{}odowska-Curie Grant Agreement No. 847648, fellowship code LCF/BQ/PI21/11830025. This work was partially funded by CEX2019-000910-S [MCIN/AEI], Fundaci\'o Cellex, Fundaci\'o Mir-Puig, and Generalitat de Catalunya through CERCA.
\bibliography{SR2}{}

\begin{thebibliography}{45}%
\makeatletter
\providecommand \@ifxundefined [1]{%
 \@ifx{#1\undefined}
}%
\providecommand \@ifnum [1]{%
 \ifnum #1\expandafter \@firstoftwo
 \else \expandafter \@secondoftwo
 \fi
}%
\providecommand \@ifx [1]{%
 \ifx #1\expandafter \@firstoftwo
 \else \expandafter \@secondoftwo
 \fi
}%
\providecommand \natexlab [1]{#1}%
\providecommand \enquote  [1]{``#1''}%
\providecommand \bibnamefont  [1]{#1}%
\providecommand \bibfnamefont [1]{#1}%
\providecommand \citenamefont [1]{#1}%
\providecommand \href@noop [0]{\@secondoftwo}%
\providecommand \href [0]{\begingroup \@sanitize@url \@href}%
\providecommand \@href[1]{\@@startlink{#1}\@@href}%
\providecommand \@@href[1]{\endgroup#1\@@endlink}%
\providecommand \@sanitize@url [0]{\catcode `\\12\catcode `\$12\catcode
  `\&12\catcode `\#12\catcode `\^12\catcode `\_12\catcode `\%12\relax}%
\providecommand \@@startlink[1]{}%
\providecommand \@@endlink[0]{}%
\providecommand \url  [0]{\begingroup\@sanitize@url \@url }%
\providecommand \@url [1]{\endgroup\@href {#1}{\urlprefix }}%
\providecommand \urlprefix  [0]{URL }%
\providecommand \Eprint [0]{\href }%
\providecommand \doibase [0]{https://doi.org/}%
\providecommand \selectlanguage [0]{\@gobble}%
\providecommand \bibinfo  [0]{\@secondoftwo}%
\providecommand \bibfield  [0]{\@secondoftwo}%
\providecommand \translation [1]{[#1]}%
\providecommand \BibitemOpen [0]{}%
\providecommand \bibitemStop [0]{}%
\providecommand \bibitemNoStop [0]{.\EOS\space}%
\providecommand \EOS [0]{\spacefactor3000\relax}%
\providecommand \BibitemShut  [1]{\csname bibitem#1\endcsname}%
\let\auto@bib@innerbib\@empty
\bibitem [{\citenamefont {Abbe}(1873)}]{Abbe}%
  \BibitemOpen
  \bibfield  {author} {\bibinfo {author} {\bibfnamefont {E.}~\bibnamefont
  {Abbe}},\ }\href {https://doi.org/10.1007/BF02956173} {\bibfield  {journal}
  {\bibinfo  {journal} {Arch. Für Mikrosk. Anat.}\ }\textbf {\bibinfo {volume}
  {9}},\ \bibinfo {pages} {413} (\bibinfo {year} {1873})}\BibitemShut {NoStop}%
\bibitem [{\citenamefont {Rayleigh}(1879)}]{Rayleigh}%
  \BibitemOpen
  \bibfield  {author} {\bibinfo {author} {\bibfnamefont {F.~R.~S.}\
  \bibnamefont {Rayleigh}},\ }\href {https://doi.org/10.1080/14786447908639684}
  {\bibfield  {journal} {\bibinfo  {journal} {Lond. Edinb. Dublin Philos. Mag.
  J. Sci.}\ }\textbf {\bibinfo {volume} {8}},\ \bibinfo {pages} {261} (\bibinfo
  {year} {1879})}\BibitemShut {NoStop}%
\bibitem [{\citenamefont {Hell}\ and\ \citenamefont
  {Wichmann}(1994)}]{Hell:94}%
  \BibitemOpen
  \bibfield  {author} {\bibinfo {author} {\bibfnamefont {S.~W.}\ \bibnamefont
  {Hell}}\ and\ \bibinfo {author} {\bibfnamefont {J.}~\bibnamefont
  {Wichmann}},\ }\href {https://doi.org/10.1364/OL.19.000780} {\bibfield
  {journal} {\bibinfo  {journal} {Opt. Lett.}\ }\textbf {\bibinfo {volume}
  {19}},\ \bibinfo {pages} {780} (\bibinfo {year} {1994})}\BibitemShut
  {NoStop}%
\bibitem [{\citenamefont {Klar}\ \emph {et~al.}(2000)\citenamefont {Klar},
  \citenamefont {Jakobs}, \citenamefont {Dyba}, \citenamefont {Egner},\ and\
  \citenamefont {Hell}}]{Klar8206}%
  \BibitemOpen
  \bibfield  {author} {\bibinfo {author} {\bibfnamefont {T.~A.}\ \bibnamefont
  {Klar}}, \bibinfo {author} {\bibfnamefont {S.}~\bibnamefont {Jakobs}},
  \bibinfo {author} {\bibfnamefont {M.}~\bibnamefont {Dyba}}, \bibinfo {author}
  {\bibfnamefont {A.}~\bibnamefont {Egner}},\ and\ \bibinfo {author}
  {\bibfnamefont {S.~W.}\ \bibnamefont {Hell}},\ }\href
  {https://doi.org/10.1073/pnas.97.15.8206} {\bibfield  {journal} {\bibinfo
  {journal} {Proceedings of the National Academy of Sciences}\ }\textbf
  {\bibinfo {volume} {97}},\ \bibinfo {pages} {8206} (\bibinfo {year}
  {2000})}\BibitemShut {NoStop}%
\bibitem [{\citenamefont {Betzig}\ \emph {et~al.}(2006)\citenamefont {Betzig},
  \citenamefont {Patterson}, \citenamefont {Sougrat}, \citenamefont
  {Lindwasser}, \citenamefont {Olenych}, \citenamefont {Bonifacino},
  \citenamefont {Davidson}, \citenamefont {Lippincott-Schwartz},\ and\
  \citenamefont {Hess}}]{Betzig1642}%
  \BibitemOpen
  \bibfield  {author} {\bibinfo {author} {\bibfnamefont {E.}~\bibnamefont
  {Betzig}}, \bibinfo {author} {\bibfnamefont {G.~H.}\ \bibnamefont
  {Patterson}}, \bibinfo {author} {\bibfnamefont {R.}~\bibnamefont {Sougrat}},
  \bibinfo {author} {\bibfnamefont {O.~W.}\ \bibnamefont {Lindwasser}},
  \bibinfo {author} {\bibfnamefont {S.}~\bibnamefont {Olenych}}, \bibinfo
  {author} {\bibfnamefont {J.~S.}\ \bibnamefont {Bonifacino}}, \bibinfo
  {author} {\bibfnamefont {M.~W.}\ \bibnamefont {Davidson}}, \bibinfo {author}
  {\bibfnamefont {J.}~\bibnamefont {Lippincott-Schwartz}},\ and\ \bibinfo
  {author} {\bibfnamefont {H.~F.}\ \bibnamefont {Hess}},\ }\href
  {https://doi.org/10.1126/science.1127344} {\bibfield  {journal} {\bibinfo
  {journal} {Science}\ }\textbf {\bibinfo {volume} {313}},\ \bibinfo {pages}
  {1642} (\bibinfo {year} {2006})}\BibitemShut {NoStop}%
\bibitem [{\citenamefont {{Helstrom}}(1973)}]{Helstrom73}%
  \BibitemOpen
  \bibfield  {author} {\bibinfo {author} {\bibfnamefont {C.}~\bibnamefont
  {{Helstrom}}},\ }\href {https://doi.org/10.1109/TIT.1973.1055052} {\bibfield
  {journal} {\bibinfo  {journal} {IEEE Transactions on Information Theory}\
  }\textbf {\bibinfo {volume} {19}},\ \bibinfo {pages} {389} (\bibinfo {year}
  {1973})}\BibitemShut {NoStop}%
\bibitem [{\citenamefont {Hsu}\ \emph {et~al.}(2004)\citenamefont {Hsu},
  \citenamefont {Delaubert}, \citenamefont {Lam},\ and\ \citenamefont
  {Bowen}}]{Hsu_2004}%
  \BibitemOpen
  \bibfield  {author} {\bibinfo {author} {\bibfnamefont {M.~T.~L.}\
  \bibnamefont {Hsu}}, \bibinfo {author} {\bibfnamefont {V.}~\bibnamefont
  {Delaubert}}, \bibinfo {author} {\bibfnamefont {P.~K.}\ \bibnamefont {Lam}},\
  and\ \bibinfo {author} {\bibfnamefont {W.~P.}\ \bibnamefont {Bowen}},\ }\href
  {https://doi.org/10.1088/1464-4266/6/12/003} {\bibfield  {journal} {\bibinfo
  {journal} {Journal of Optics B: Quantum and Semiclassical Optics}\ }\textbf
  {\bibinfo {volume} {6}},\ \bibinfo {pages} {495} (\bibinfo {year}
  {2004})}\BibitemShut {NoStop}%
\bibitem [{\citenamefont {Delaubert}\ \emph {et~al.}(2008)\citenamefont
  {Delaubert}, \citenamefont {Treps}, \citenamefont {Fabre}, \citenamefont
  {Bachor},\ and\ \citenamefont {R{\'{e}}fr{\'{e}}gier}}]{Delaubert_2008}%
  \BibitemOpen
  \bibfield  {author} {\bibinfo {author} {\bibfnamefont {V.}~\bibnamefont
  {Delaubert}}, \bibinfo {author} {\bibfnamefont {N.}~\bibnamefont {Treps}},
  \bibinfo {author} {\bibfnamefont {C.}~\bibnamefont {Fabre}}, \bibinfo
  {author} {\bibfnamefont {H.~A.}\ \bibnamefont {Bachor}},\ and\ \bibinfo
  {author} {\bibfnamefont {P.}~\bibnamefont {R{\'{e}}fr{\'{e}}gier}},\ }\href
  {https://doi.org/10.1209/0295-5075/81/44001} {\bibfield  {journal} {\bibinfo
  {journal} {{EPL} (Europhysics Letters)}\ }\textbf {\bibinfo {volume} {81}},\
  \bibinfo {pages} {44001} (\bibinfo {year} {2008})}\BibitemShut {NoStop}%
\bibitem [{\citenamefont {Tsang}(2019)}]{Tsang_review}%
  \BibitemOpen
  \bibfield  {author} {\bibinfo {author} {\bibfnamefont {M.}~\bibnamefont
  {Tsang}},\ }\href {https://doi.org/10.1080/00107514.2020.1736375} {\bibfield
  {journal} {\bibinfo  {journal} {Contemporary Physics}\ }\textbf {\bibinfo
  {volume} {60}},\ \bibinfo {pages} {279} (\bibinfo {year} {2019})}\BibitemShut
  {NoStop}%
\bibitem [{\citenamefont {Helstrom}(1976)}]{helstrom1976}%
  \BibitemOpen
  \bibfield  {author} {\bibinfo {author} {\bibfnamefont {C.~W.}\ \bibnamefont
  {Helstrom}},\ }\href@noop {} {\emph {\bibinfo {title} {Quantum detection and
  estimation theory}}},\ Vol.~\bibinfo {volume} {3}\ (\bibinfo  {publisher}
  {Academic press, New York},\ \bibinfo {year} {1976})\BibitemShut {NoStop}%
\bibitem [{\citenamefont {Holevo}(2011)}]{holevo2011probabilistic}%
  \BibitemOpen
  \bibfield  {author} {\bibinfo {author} {\bibfnamefont {A.~S.}\ \bibnamefont
  {Holevo}},\ }\href@noop {} {\emph {\bibinfo {title} {Probabilistic and
  statistical aspects of quantum theory}}},\ Vol.~\bibinfo {volume} {1}\
  (\bibinfo  {publisher} {Springer, Berlin},\ \bibinfo {year}
  {2011})\BibitemShut {NoStop}%
\bibitem [{\citenamefont {Giovannetti}\ \emph {et~al.}(2011)\citenamefont
  {Giovannetti}, \citenamefont {Lloyd},\ and\ \citenamefont
  {Maccone}}]{GiovannettiLoydMaccone}%
  \BibitemOpen
  \bibfield  {author} {\bibinfo {author} {\bibfnamefont {V.}~\bibnamefont
  {Giovannetti}}, \bibinfo {author} {\bibfnamefont {S.}~\bibnamefont {Lloyd}},\
  and\ \bibinfo {author} {\bibfnamefont {L.}~\bibnamefont {Maccone}},\ }\href
  {https://doi.org/10.1038/nphoton.2011.35} {\bibfield  {journal} {\bibinfo
  {journal} {Nature Photonics}\ }\textbf {\bibinfo {volume} {5}},\ \bibinfo
  {pages} {222} (\bibinfo {year} {2011})}\BibitemShut {NoStop}%
\bibitem [{\citenamefont {Paris}(2009)}]{Paris2009}%
  \BibitemOpen
  \bibfield  {author} {\bibinfo {author} {\bibfnamefont {M.~G.~A.}\
  \bibnamefont {Paris}},\ }\href {https://doi.org/10.1142/S0219749909004839}
  {\bibfield  {journal} {\bibinfo  {journal} {International Journal of Quantum
  Information}\ }\textbf {\bibinfo {volume} {07}},\ \bibinfo {pages} {125}
  (\bibinfo {year} {2009})}\BibitemShut {NoStop}%
\bibitem [{\citenamefont {Pezzè}\ and\ \citenamefont
  {Smerzi}(2014)}]{Luca_Augusto_review}%
  \BibitemOpen
  \bibfield  {author} {\bibinfo {author} {\bibfnamefont {L.}~\bibnamefont
  {Pezzè}}\ and\ \bibinfo {author} {\bibfnamefont {A.}~\bibnamefont
  {Smerzi}},\ }in\ \href {https://doi.org/10.3254/978-1-61499-448-0-691} {\emph
  {\bibinfo {booktitle} {Proceedings of the International School of Physics
  "Enrico Fermi"}}},\ \bibinfo {series and number} {\bibinfo {number} {Course
  188, Varenna}},\ \bibinfo {editor} {edited by\ \bibinfo {editor}
  {\bibfnamefont {G.~M.}\ \bibnamefont {Tino}}\ and\ \bibinfo {editor}
  {\bibfnamefont {M.~A.}\ \bibnamefont {Kasevich}}}\ (\bibinfo  {publisher}
  {IOS Press, Amsterdam},\ \bibinfo {year} {2014})\ pp.\ \bibinfo {pages} {691
  -- 741}\BibitemShut {NoStop}%
\bibitem [{\citenamefont {Tsang}\ \emph {et~al.}(2016)\citenamefont {Tsang},
  \citenamefont {Nair},\ and\ \citenamefont {Lu}}]{Tsang_PRX}%
  \BibitemOpen
  \bibfield  {author} {\bibinfo {author} {\bibfnamefont {M.}~\bibnamefont
  {Tsang}}, \bibinfo {author} {\bibfnamefont {R.}~\bibnamefont {Nair}},\ and\
  \bibinfo {author} {\bibfnamefont {X.-M.}\ \bibnamefont {Lu}},\ }\href
  {https://doi.org/10.1103/PhysRevX.6.031033} {\bibfield  {journal} {\bibinfo
  {journal} {Phys. Rev. X}\ }\textbf {\bibinfo {volume} {6}},\ \bibinfo {pages}
  {031033} (\bibinfo {year} {2016})}\BibitemShut {NoStop}%
\bibitem [{\citenamefont {Rehacek}\ \emph {et~al.}(2017)\citenamefont
  {Rehacek}, \citenamefont {Pa\'{u}r}, \citenamefont {Stoklasa}, \citenamefont
  {Hradil},\ and\ \citenamefont {S\'{a}nchez-Soto}}]{Rehacek:17}%
  \BibitemOpen
  \bibfield  {author} {\bibinfo {author} {\bibfnamefont {J.}~\bibnamefont
  {Rehacek}}, \bibinfo {author} {\bibfnamefont {M.}~\bibnamefont {Pa\'{u}r}},
  \bibinfo {author} {\bibfnamefont {B.}~\bibnamefont {Stoklasa}}, \bibinfo
  {author} {\bibfnamefont {Z.}~\bibnamefont {Hradil}},\ and\ \bibinfo {author}
  {\bibfnamefont {L.~L.}\ \bibnamefont {S\'{a}nchez-Soto}},\ }\href
  {https://doi.org/10.1364/OL.42.000231} {\bibfield  {journal} {\bibinfo
  {journal} {Opt. Lett.}\ }\textbf {\bibinfo {volume} {42}},\ \bibinfo {pages}
  {231} (\bibinfo {year} {2017})}\BibitemShut {NoStop}%
\bibitem [{\citenamefont {Pa\'{u}r}\ \emph {et~al.}(2016)\citenamefont
  {Pa\'{u}r}, \citenamefont {Stoklasa}, \citenamefont {Hradil}, \citenamefont
  {S\'{a}nchez-Soto},\ and\ \citenamefont {Rehacek}}]{Paur:16}%
  \BibitemOpen
  \bibfield  {author} {\bibinfo {author} {\bibfnamefont {M.}~\bibnamefont
  {Pa\'{u}r}}, \bibinfo {author} {\bibfnamefont {B.}~\bibnamefont {Stoklasa}},
  \bibinfo {author} {\bibfnamefont {Z.}~\bibnamefont {Hradil}}, \bibinfo
  {author} {\bibfnamefont {L.~L.}\ \bibnamefont {S\'{a}nchez-Soto}},\ and\
  \bibinfo {author} {\bibfnamefont {J.}~\bibnamefont {Rehacek}},\ }\href
  {https://doi.org/10.1364/OPTICA.3.001144} {\bibfield  {journal} {\bibinfo
  {journal} {Optica}\ }\textbf {\bibinfo {volume} {3}},\ \bibinfo {pages}
  {1144} (\bibinfo {year} {2016})}\BibitemShut {NoStop}%
\bibitem [{\citenamefont {Tang}\ \emph {et~al.}(2016)\citenamefont {Tang},
  \citenamefont {Durak},\ and\ \citenamefont {Ling}}]{Tang:16}%
  \BibitemOpen
  \bibfield  {author} {\bibinfo {author} {\bibfnamefont {Z.~S.}\ \bibnamefont
  {Tang}}, \bibinfo {author} {\bibfnamefont {K.}~\bibnamefont {Durak}},\ and\
  \bibinfo {author} {\bibfnamefont {A.}~\bibnamefont {Ling}},\ }\href
  {https://doi.org/10.1364/OE.24.022004} {\bibfield  {journal} {\bibinfo
  {journal} {Opt. Express}\ }\textbf {\bibinfo {volume} {24}},\ \bibinfo
  {pages} {22004} (\bibinfo {year} {2016})}\BibitemShut {NoStop}%
\bibitem [{\citenamefont {Yang}\ \emph {et~al.}(2016)\citenamefont {Yang},
  \citenamefont {Tashchilina}, \citenamefont {Moiseev}, \citenamefont {Simon},\
  and\ \citenamefont {Lvovsky}}]{Yang:16}%
  \BibitemOpen
  \bibfield  {author} {\bibinfo {author} {\bibfnamefont {F.}~\bibnamefont
  {Yang}}, \bibinfo {author} {\bibfnamefont {A.}~\bibnamefont {Tashchilina}},
  \bibinfo {author} {\bibfnamefont {E.~S.}\ \bibnamefont {Moiseev}}, \bibinfo
  {author} {\bibfnamefont {C.}~\bibnamefont {Simon}},\ and\ \bibinfo {author}
  {\bibfnamefont {A.~I.}\ \bibnamefont {Lvovsky}},\ }\href
  {https://doi.org/10.1364/OPTICA.3.001148} {\bibfield  {journal} {\bibinfo
  {journal} {Optica}\ }\textbf {\bibinfo {volume} {3}},\ \bibinfo {pages}
  {1148} (\bibinfo {year} {2016})}\BibitemShut {NoStop}%
\bibitem [{\citenamefont {Tham}\ \emph {et~al.}(2017)\citenamefont {Tham},
  \citenamefont {Ferretti},\ and\ \citenamefont {Steinberg}}]{Tham:2017}%
  \BibitemOpen
  \bibfield  {author} {\bibinfo {author} {\bibfnamefont {W.-K.}\ \bibnamefont
  {Tham}}, \bibinfo {author} {\bibfnamefont {H.}~\bibnamefont {Ferretti}},\
  and\ \bibinfo {author} {\bibfnamefont {A.~M.}\ \bibnamefont {Steinberg}},\
  }\href {https://doi.org/10.1103/PhysRevLett.118.070801} {\bibfield  {journal}
  {\bibinfo  {journal} {Phys. Rev. Lett.}\ }\textbf {\bibinfo {volume} {118}},\
  \bibinfo {pages} {070801} (\bibinfo {year} {2017})}\BibitemShut {NoStop}%
\bibitem [{\citenamefont {Boucher}\ \emph {et~al.}(2020)\citenamefont
  {Boucher}, \citenamefont {Fabre}, \citenamefont {Labroille},\ and\
  \citenamefont {Treps}}]{Boucher:20}%
  \BibitemOpen
  \bibfield  {author} {\bibinfo {author} {\bibfnamefont {P.}~\bibnamefont
  {Boucher}}, \bibinfo {author} {\bibfnamefont {C.}~\bibnamefont {Fabre}},
  \bibinfo {author} {\bibfnamefont {G.}~\bibnamefont {Labroille}},\ and\
  \bibinfo {author} {\bibfnamefont {N.}~\bibnamefont {Treps}},\ }\href
  {https://doi.org/10.1364/OPTICA.404746} {\bibfield  {journal} {\bibinfo
  {journal} {Optica}\ }\textbf {\bibinfo {volume} {7}},\ \bibinfo {pages}
  {1621} (\bibinfo {year} {2020})}\BibitemShut {NoStop}%
\bibitem [{\citenamefont {Zanforlin}\ \emph {et~al.}(2022)\citenamefont
  {Zanforlin}, \citenamefont {Lupo}, \citenamefont {Connolly}, \citenamefont
  {Kok}, \citenamefont {Buller},\ and\ \citenamefont {Huang}}]{Zanforlin:2022}%
  \BibitemOpen
  \bibfield  {author} {\bibinfo {author} {\bibfnamefont {U.}~\bibnamefont
  {Zanforlin}}, \bibinfo {author} {\bibfnamefont {C.}~\bibnamefont {Lupo}},
  \bibinfo {author} {\bibfnamefont {P.~W.~R.}\ \bibnamefont {Connolly}},
  \bibinfo {author} {\bibfnamefont {P.}~\bibnamefont {Kok}}, \bibinfo {author}
  {\bibfnamefont {G.~S.}\ \bibnamefont {Buller}},\ and\ \bibinfo {author}
  {\bibfnamefont {Z.}~\bibnamefont {Huang}},\ }\href@noop {} {\bibfield
  {journal} {\bibinfo  {journal} {arXiv:2202.09406}\ } (\bibinfo {year}
  {2022})}\BibitemShut {NoStop}%
\bibitem [{\citenamefont {Nair}\ and\ \citenamefont {Tsang}(2016)}]{Nair_2016}%
  \BibitemOpen
  \bibfield  {author} {\bibinfo {author} {\bibfnamefont {R.}~\bibnamefont
  {Nair}}\ and\ \bibinfo {author} {\bibfnamefont {M.}~\bibnamefont {Tsang}},\
  }\href {https://doi.org/10.1103/PhysRevLett.117.190801} {\bibfield  {journal}
  {\bibinfo  {journal} {Phys. Rev. Lett.}\ }\textbf {\bibinfo {volume} {117}},\
  \bibinfo {pages} {190801} (\bibinfo {year} {2016})}\BibitemShut {NoStop}%
\bibitem [{\citenamefont {Lupo}\ and\ \citenamefont
  {Pirandola}(2016)}]{LupoPirandola}%
  \BibitemOpen
  \bibfield  {author} {\bibinfo {author} {\bibfnamefont {C.}~\bibnamefont
  {Lupo}}\ and\ \bibinfo {author} {\bibfnamefont {S.}~\bibnamefont
  {Pirandola}},\ }\href {https://doi.org/10.1103/PhysRevLett.117.190802}
  {\bibfield  {journal} {\bibinfo  {journal} {Phys. Rev. Lett.}\ }\textbf
  {\bibinfo {volume} {117}},\ \bibinfo {pages} {190802} (\bibinfo {year}
  {2016})}\BibitemShut {NoStop}%
\bibitem [{\citenamefont {Larson}\ and\ \citenamefont
  {Saleh}(2018)}]{Larson:18}%
  \BibitemOpen
  \bibfield  {author} {\bibinfo {author} {\bibfnamefont {W.}~\bibnamefont
  {Larson}}\ and\ \bibinfo {author} {\bibfnamefont {B.~E.~A.}\ \bibnamefont
  {Saleh}},\ }\href {https://doi.org/10.1364/OPTICA.5.001382} {\bibfield
  {journal} {\bibinfo  {journal} {Optica}\ }\textbf {\bibinfo {volume} {5}},\
  \bibinfo {pages} {1382} (\bibinfo {year} {2018})}\BibitemShut {NoStop}%
\bibitem [{\citenamefont {Tsang}\ and\ \citenamefont
  {Nair}(2019)}]{tsang2019comment}%
  \BibitemOpen
  \bibfield  {author} {\bibinfo {author} {\bibfnamefont {M.}~\bibnamefont
  {Tsang}}\ and\ \bibinfo {author} {\bibfnamefont {R.}~\bibnamefont {Nair}},\
  }\href {https://doi.org/10.1364/OPTICA.6.000400} {\bibfield  {journal}
  {\bibinfo  {journal} {Optica}\ }\textbf {\bibinfo {volume} {6}},\ \bibinfo
  {pages} {400} (\bibinfo {year} {2019})}\BibitemShut {NoStop}%
\bibitem [{\citenamefont {Larson}\ and\ \citenamefont
  {Saleh}(2019)}]{Larson19reply}%
  \BibitemOpen
  \bibfield  {author} {\bibinfo {author} {\bibfnamefont {W.}~\bibnamefont
  {Larson}}\ and\ \bibinfo {author} {\bibfnamefont {B.~E.~A.}\ \bibnamefont
  {Saleh}},\ }\href {https://doi.org/10.1364/OPTICA.6.000402} {\bibfield
  {journal} {\bibinfo  {journal} {Optica}\ }\textbf {\bibinfo {volume} {6}},\
  \bibinfo {pages} {402} (\bibinfo {year} {2019})}\BibitemShut {NoStop}%
\bibitem [{\citenamefont {Hradil}\ \emph {et~al.}(2019)\citenamefont {Hradil},
  \citenamefont {\v{R}eh\'{a}\v{c}ek}, \citenamefont {S\'{a}nchez-Soto},\ and\
  \citenamefont {Englert}}]{Hradil19QFI}%
  \BibitemOpen
  \bibfield  {author} {\bibinfo {author} {\bibfnamefont {Z.}~\bibnamefont
  {Hradil}}, \bibinfo {author} {\bibfnamefont {J.}~\bibnamefont
  {\v{R}eh\'{a}\v{c}ek}}, \bibinfo {author} {\bibfnamefont {L.}~\bibnamefont
  {S\'{a}nchez-Soto}},\ and\ \bibinfo {author} {\bibfnamefont {B.-G.}\
  \bibnamefont {Englert}},\ }\href {https://doi.org/10.1364/OPTICA.6.001437}
  {\bibfield  {journal} {\bibinfo  {journal} {Optica}\ }\textbf {\bibinfo
  {volume} {6}},\ \bibinfo {pages} {1437} (\bibinfo {year} {2019})}\BibitemShut
  {NoStop}%
\bibitem [{\citenamefont {Hradil}\ \emph {et~al.}(2021)\citenamefont {Hradil},
  \citenamefont {Koutn\'{y}},\ and\ \citenamefont
  {\v{R}eh\'{a}\v{c}ek}}]{Hradil21Partial}%
  \BibitemOpen
  \bibfield  {author} {\bibinfo {author} {\bibfnamefont {Z.}~\bibnamefont
  {Hradil}}, \bibinfo {author} {\bibfnamefont {D.}~\bibnamefont {Koutn\'{y}}},\
  and\ \bibinfo {author} {\bibfnamefont {J.}~\bibnamefont
  {\v{R}eh\'{a}\v{c}ek}},\ }\href {https://doi.org/10.1364/OL.417988}
  {\bibfield  {journal} {\bibinfo  {journal} {Opt. Lett.}\ }\textbf {\bibinfo
  {volume} {46}},\ \bibinfo {pages} {1728} (\bibinfo {year}
  {2021})}\BibitemShut {NoStop}%
\bibitem [{\citenamefont {Kurdzialek}(2021)}]{kurdzialek2021sources}%
  \BibitemOpen
  \bibfield  {author} {\bibinfo {author} {\bibfnamefont {S.}~\bibnamefont
  {Kurdzialek}},\ }\href@noop {} {\bibfield  {journal} {\bibinfo  {journal}
  {arXiv:2103.12096}\ } (\bibinfo {year} {2021})}\BibitemShut {NoStop}%
\bibitem [{\citenamefont {Liang}\ \emph {et~al.}(2021)\citenamefont {Liang},
  \citenamefont {Wadood},\ and\ \citenamefont {Vamivakas}}]{Liang:21}%
  \BibitemOpen
  \bibfield  {author} {\bibinfo {author} {\bibfnamefont {K.}~\bibnamefont
  {Liang}}, \bibinfo {author} {\bibfnamefont {S.~A.}\ \bibnamefont {Wadood}},\
  and\ \bibinfo {author} {\bibfnamefont {A.~N.}\ \bibnamefont {Vamivakas}},\
  }\href {https://doi.org/10.1364/OPTICA.403497} {\bibfield  {journal}
  {\bibinfo  {journal} {Optica}\ }\textbf {\bibinfo {volume} {8}},\ \bibinfo
  {pages} {243} (\bibinfo {year} {2021})}\BibitemShut {NoStop}%
\bibitem [{\citenamefont {Wadood}\ \emph {et~al.}(2021)\citenamefont {Wadood},
  \citenamefont {Liang}, \citenamefont {Zhou}, \citenamefont {Yang},
  \citenamefont {Alonso}, \citenamefont {Qian}, \citenamefont {Malhotra},
  \citenamefont {Rafsanjani}, \citenamefont {Jordan}, \citenamefont {Boyd},\
  and\ \citenamefont {Vamivakas}}]{Wadood:21}%
  \BibitemOpen
  \bibfield  {author} {\bibinfo {author} {\bibfnamefont {S.~A.}\ \bibnamefont
  {Wadood}}, \bibinfo {author} {\bibfnamefont {K.}~\bibnamefont {Liang}},
  \bibinfo {author} {\bibfnamefont {Y.}~\bibnamefont {Zhou}}, \bibinfo {author}
  {\bibfnamefont {J.}~\bibnamefont {Yang}}, \bibinfo {author} {\bibfnamefont
  {M.~A.}\ \bibnamefont {Alonso}}, \bibinfo {author} {\bibfnamefont {X.-F.}\
  \bibnamefont {Qian}}, \bibinfo {author} {\bibfnamefont {T.}~\bibnamefont
  {Malhotra}}, \bibinfo {author} {\bibfnamefont {S.~M.~H.}\ \bibnamefont
  {Rafsanjani}}, \bibinfo {author} {\bibfnamefont {A.~N.}\ \bibnamefont
  {Jordan}}, \bibinfo {author} {\bibfnamefont {R.~W.}\ \bibnamefont {Boyd}},\
  and\ \bibinfo {author} {\bibfnamefont {A.~N.}\ \bibnamefont {Vamivakas}},\
  }\href {https://doi.org/10.1364/OE.427734} {\bibfield  {journal} {\bibinfo
  {journal} {Opt. Express}\ }\textbf {\bibinfo {volume} {29}},\ \bibinfo
  {pages} {22034} (\bibinfo {year} {2021})}\BibitemShut {NoStop}%
\bibitem [{\citenamefont {De}\ \emph {et~al.}(2021)\citenamefont {De},
  \citenamefont {Gil-Lopez}, \citenamefont {Brecht}, \citenamefont
  {Silberhorn}, \citenamefont {S\'anchez-Soto}, \citenamefont {Hradil},\ and\
  \citenamefont {\ifmmode \check{R}\else \v{R}\fi{}eh\'a\ifmmode~\check{c}\else
  \v{c}\fi{}ek}}]{De:2021}%
  \BibitemOpen
  \bibfield  {author} {\bibinfo {author} {\bibfnamefont {S.}~\bibnamefont
  {De}}, \bibinfo {author} {\bibfnamefont {J.}~\bibnamefont {Gil-Lopez}},
  \bibinfo {author} {\bibfnamefont {B.}~\bibnamefont {Brecht}}, \bibinfo
  {author} {\bibfnamefont {C.}~\bibnamefont {Silberhorn}}, \bibinfo {author}
  {\bibfnamefont {L.~L.}\ \bibnamefont {S\'anchez-Soto}}, \bibinfo {author}
  {\bibfnamefont {Z.~c.~v.}\ \bibnamefont {Hradil}},\ and\ \bibinfo {author}
  {\bibfnamefont {J.}~\bibnamefont {\ifmmode \check{R}\else
  \v{R}\fi{}eh\'a\ifmmode~\check{c}\else \v{c}\fi{}ek}},\ }\href
  {https://doi.org/10.1103/PhysRevResearch.3.033082} {\bibfield  {journal}
  {\bibinfo  {journal} {Phys. Rev. Research}\ }\textbf {\bibinfo {volume}
  {3}},\ \bibinfo {pages} {033082} (\bibinfo {year} {2021})}\BibitemShut
  {NoStop}%
\bibitem [{\citenamefont {Tsang}(2021)}]{Tsang2021poissonquantum}%
  \BibitemOpen
  \bibfield  {author} {\bibinfo {author} {\bibfnamefont {M.}~\bibnamefont
  {Tsang}},\ }\href {https://doi.org/10.22331/q-2021-08-19-527} {\bibfield
  {journal} {\bibinfo  {journal} {{Quantum}}\ }\textbf {\bibinfo {volume}
  {5}},\ \bibinfo {pages} {527} (\bibinfo {year} {2021})}\BibitemShut {NoStop}%
\bibitem [{\citenamefont {Holevo}(1975)}]{Holevo:1975}%
  \BibitemOpen
  \bibfield  {author} {\bibinfo {author} {\bibfnamefont {A.}~\bibnamefont
  {Holevo}},\ }\href {https://doi.org/10.1109/TIT.1975.1055441} {\bibfield
  {journal} {\bibinfo  {journal} {IEEE Transactions on Information Theory}\
  }\textbf {\bibinfo {volume} {21}},\ \bibinfo {pages} {533} (\bibinfo {year}
  {1975})}\BibitemShut {NoStop}%
\bibitem [{\citenamefont {Weedbrook}\ \emph {et~al.}(2012)\citenamefont
  {Weedbrook}, \citenamefont {Pirandola}, \citenamefont
  {Garc{\'\i}a-Patr{\'o}n}, \citenamefont {Cerf}, \citenamefont {Ralph},
  \citenamefont {Shapiro},\ and\ \citenamefont {Lloyd}}]{Weedbrook:2012}%
  \BibitemOpen
  \bibfield  {author} {\bibinfo {author} {\bibfnamefont {C.}~\bibnamefont
  {Weedbrook}}, \bibinfo {author} {\bibfnamefont {S.}~\bibnamefont
  {Pirandola}}, \bibinfo {author} {\bibfnamefont {R.}~\bibnamefont
  {Garc{\'\i}a-Patr{\'o}n}}, \bibinfo {author} {\bibfnamefont {N.~J.}\
  \bibnamefont {Cerf}}, \bibinfo {author} {\bibfnamefont {T.~C.}\ \bibnamefont
  {Ralph}}, \bibinfo {author} {\bibfnamefont {J.~H.}\ \bibnamefont {Shapiro}},\
  and\ \bibinfo {author} {\bibfnamefont {S.}~\bibnamefont {Lloyd}},\
  }\href@noop {} {\bibfield  {journal} {\bibinfo  {journal} {Reviews of Modern
  Physics}\ }\textbf {\bibinfo {volume} {84}},\ \bibinfo {pages} {621}
  (\bibinfo {year} {2012})}\BibitemShut {NoStop}%
\bibitem [{\citenamefont {Adesso}\ \emph {et~al.}(2014)\citenamefont {Adesso},
  \citenamefont {Ragy},\ and\ \citenamefont {Lee}}]{Adesso:2014}%
  \BibitemOpen
  \bibfield  {author} {\bibinfo {author} {\bibfnamefont {G.}~\bibnamefont
  {Adesso}}, \bibinfo {author} {\bibfnamefont {S.}~\bibnamefont {Ragy}},\ and\
  \bibinfo {author} {\bibfnamefont {A.~R.}\ \bibnamefont {Lee}},\ }\href@noop
  {} {\bibfield  {journal} {\bibinfo  {journal} {Open Systems \& Information
  Dynamics}\ }\textbf {\bibinfo {volume} {21}},\ \bibinfo {pages} {1440001}
  (\bibinfo {year} {2014})}\BibitemShut {NoStop}%
\bibitem [{\citenamefont {Sorelli}\ \emph {et~al.}(2022)\citenamefont
  {Sorelli}, \citenamefont {Gessner}, \citenamefont {Walschaers},\ and\
  \citenamefont {Treps}}]{Sorelli:2022}%
  \BibitemOpen
  \bibfield  {author} {\bibinfo {author} {\bibfnamefont {G.}~\bibnamefont
  {Sorelli}}, \bibinfo {author} {\bibfnamefont {M.}~\bibnamefont {Gessner}},
  \bibinfo {author} {\bibfnamefont {M.}~\bibnamefont {Walschaers}},\ and\
  \bibinfo {author} {\bibfnamefont {N.}~\bibnamefont {Treps}},\ }\href@noop {}
  {\bibfield  {journal} {\bibinfo  {journal} {arXiv:2202.10355}\ } (\bibinfo
  {year} {2022})}\BibitemShut {NoStop}%
\bibitem [{Note1()}]{Note1}%
  \BibitemOpen
  \bibinfo {note} {The matrices $A^{(l)}_{jk}$ are zero everywhere except in
  the $jk$ block where they are given by $i\sigma _y/\protect \sqrt {2}$ (for
  $l=0$), $\sigma _z/\protect \sqrt {2}$ (for $l=1$), $\protect \mathds
  {1}_2/\protect \sqrt {2}$ (for $l=2$) and $\sigma _x/\protect \sqrt {2}$ (for
  $l=3$), with $\sigma _{j=x,y,z}$ the standard Pauli matrices}\BibitemShut
  {NoStop}%
\bibitem [{\citenamefont {Mandel}\ and\ \citenamefont
  {Wolf}(1995)}]{mandel_wolf}%
  \BibitemOpen
  \bibfield  {author} {\bibinfo {author} {\bibfnamefont {L.}~\bibnamefont
  {Mandel}}\ and\ \bibinfo {author} {\bibfnamefont {E.}~\bibnamefont {Wolf}},\
  }\href@noop {} {\emph {\bibinfo {title} {Optical coherence and quantum
  optics}}}\ (\bibinfo  {publisher} {Cambridge university press, New York},\
  \bibinfo {year} {1995})\BibitemShut {NoStop}%
\bibitem [{\citenamefont {Klauder}\ and\ \citenamefont
  {Sudarshan}(2006)}]{klauder_sudarshan}%
  \BibitemOpen
  \bibfield  {author} {\bibinfo {author} {\bibfnamefont {J.~R.}\ \bibnamefont
  {Klauder}}\ and\ \bibinfo {author} {\bibfnamefont {E.~C.~G.}\ \bibnamefont
  {Sudarshan}},\ }\href@noop {} {\emph {\bibinfo {title} {Fundamentals of
  quantum optics}}}\ (\bibinfo  {publisher} {Dover Publication, New York},\
  \bibinfo {year} {2006})\BibitemShut {NoStop}%
\bibitem [{Note2()}]{Note2}%
  \BibitemOpen
  \bibinfo {note} {Here, without loss of generality, we assumed the first
  source to be displaced along the $q-$axis of phase space.}\BibitemShut
  {Stop}%
\bibitem [{\citenamefont {Karuseichyk}\ \emph {et~al.}(2021)\citenamefont
  {Karuseichyk}, \citenamefont {Sorelli}, \citenamefont {Gessner},
  \citenamefont {Walschaers},\ and\ \citenamefont {Treps}}]{karuseichyk2021}%
  \BibitemOpen
  \bibfield  {author} {\bibinfo {author} {\bibfnamefont {I.}~\bibnamefont
  {Karuseichyk}}, \bibinfo {author} {\bibfnamefont {G.}~\bibnamefont
  {Sorelli}}, \bibinfo {author} {\bibfnamefont {M.}~\bibnamefont {Gessner}},
  \bibinfo {author} {\bibfnamefont {M.}~\bibnamefont {Walschaers}},\ and\
  \bibinfo {author} {\bibfnamefont {N.}~\bibnamefont {Treps}},\ }\href@noop {}
  {\bibfield  {journal} {\bibinfo  {journal} {arXiv:2111.02233}\ } (\bibinfo
  {year} {2021})}\BibitemShut {NoStop}%
\bibitem [{\citenamefont {Serafini}(2017)}]{serafini_book}%
  \BibitemOpen
  \bibfield  {author} {\bibinfo {author} {\bibfnamefont {A.}~\bibnamefont
  {Serafini}},\ }\href@noop {} {\emph {\bibinfo {title} {Quantum continuous
  variables: a primer of theoretical methods}}}\ (\bibinfo  {publisher} {Tayor
  \& Francis, Milton Park},\ \bibinfo {year} {2017})\BibitemShut {NoStop}%
\bibitem [{\citenamefont {Gessner}\ \emph {et~al.}(2020)\citenamefont
  {Gessner}, \citenamefont {Fabre},\ and\ \citenamefont {Treps}}]{gessner2020}%
  \BibitemOpen
  \bibfield  {author} {\bibinfo {author} {\bibfnamefont {M.}~\bibnamefont
  {Gessner}}, \bibinfo {author} {\bibfnamefont {C.}~\bibnamefont {Fabre}},\
  and\ \bibinfo {author} {\bibfnamefont {N.}~\bibnamefont {Treps}},\ }\href
  {https://doi.org/10.1103/PhysRevLett.125.100501} {\bibfield  {journal}
  {\bibinfo  {journal} {Phys. Rev. Lett.}\ }\textbf {\bibinfo {volume} {125}},\
  \bibinfo {pages} {100501} (\bibinfo {year} {2020})}\BibitemShut {NoStop}%
\end{thebibliography}%
\bibliographystyle{apsrev4-2}
\end{document}